\documentclass[epj]{webofc}
\usepackage[varg]{txfonts}   
\woctitle{MESON2016 - the 14$^\textrm{th}$ International Workshop on Meson Production, Properties and Interaction}
\begin{document}
\selectlanguage{english}
\title{Electromagnetic effects on meson production: a new tool for studying the space-time evolution of heavy ion collisions}

\author{Andrzej Rybicki\inst{1}\fnsep\thanks{\email{andrzej.rybicki@ifj.edu.pl}} \and
        Antoni Szczurek\inst{1,2} \and
        Mariola K\l{}usek-Gawenda\inst{1} \and
        Nikolaos Davis\inst{1} \and
        Vitalii Ozvenchuk\inst{1} \and
        Miros\l{}aw Kie\l{}bowicz\inst{1} 
}

\institute{H.~Niewodnicza\'nski Institute of Nuclear Physics, Polish Academy of Sciences, Radzikowskiego 152, 31-342 Krak\'ow, Poland
\and
           University of Rzesz\'ow, Rejtana 16, 35-959 Rzesz\'ow, 
Poland
          }

\abstract{
We review our studies of spectator-induced electromagnetic (EM) effects on the emission of charged mesons in the final state of ultrarelativistic heavy ion collisions. We argue that these effects offer sensitivity to the distance $d_E$ between the charged meson formation zone at freeze-out and the spectator system. As such, they can serve as an independent, new tool to probe the space-time and longitudinal evolution of the system created in the collision. As a phenomenological application for this tool in the context of resonance production and decay, we obtain a first estimate of the 
time of pion emission 
from
EM effects. This we compare to existing HBT data. 
}
\maketitle
\section{Introduction}
\label{intro}
In the following we address the issue of final state electromagnetic (EM) interactions between the charged mesons emitted in high energy heavy ion collisions and the positively charged nuclear remnant (``spectator system'') which does not participate directly in the collision. We will argue that the latter effects can be used as a new tool to study the space-time features of the reaction, applicable over a wide range of collision centrality.

\section{EM effects on charged meson spectra and directed flow}
\label{emeff}

\begin{figure}[ht]
\centering
\includegraphics[width=14cm,clip]{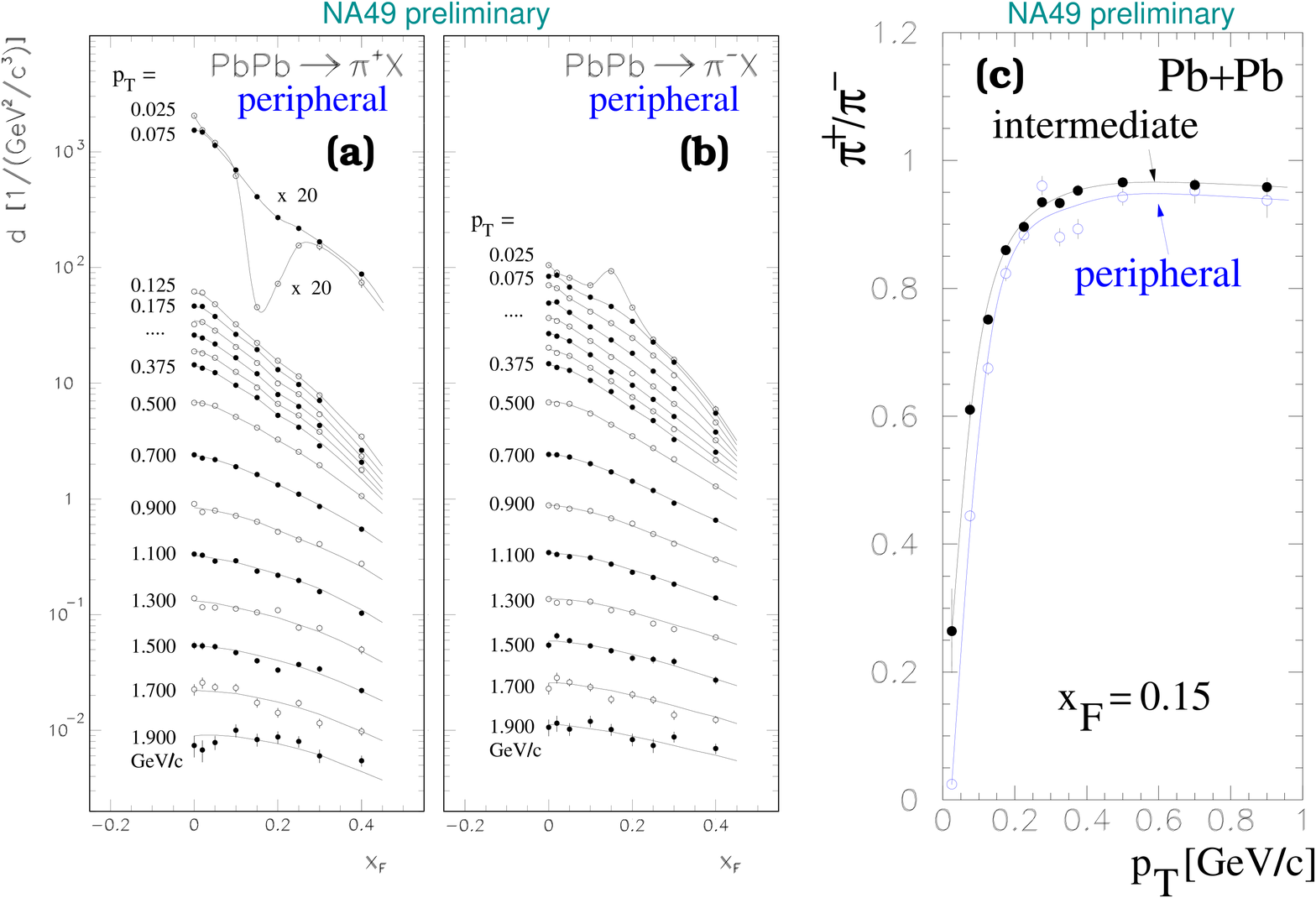}
\caption{Spectator-induced EM effects on final state charged $\pi$ meson spectra and $\pi^+/\pi^-$ ratios in Pb+Pb collisions at $\sqrt{s_{NN}}=17.3$~GeV. Double differential $d=E\frac{d^3N}{dp^3}$ distributions of (a) positive and (b) negative pions drawn as a function of $x_F=\frac{2p_L}{\sqrt{s_{NN}}}$ and transverse momentum $p_T$ (both quantities in the collision c.m.s.), and (c) ratio of $\pi^+$ over $\pi^-$ distributions at $x_F=0.15$.}
\label{fig-1}       
\end{figure}

An overview of the influence of the EM field from the spectator system over the phase space available for $\pi^+$ and $\pi^-$ meson production is presented in
Fig.~\ref{fig-1}. The measurements come from the NA49 experiment at the CERN SPS, for peripheral and intermediate centrality Pb+Pb collisions at $\sqrt{s_{NN}}=17.3$~GeV. The invariant double differential spectra of positive (negative) pions (panels a,b) display an evident strong depletion (enhancement) at low transverse momenta in the vicinity of $x_F=0.15=m_\pi/m_p$, which corresponds to pions moving parallel to the spectator system and with the same velocity (rapidity). This apparent repulsion of positive (attraction of negative) pions from spectator vicinity results in a very large, isospin symmetry-breaking distortion of $\pi^+/\pi^-$ ratios as a function of pion transverse momentum as apparent in panel (c). What may, at first glance, appear surprising is that the centrality dependence of this effect is quite weak at least between peripheral and intermediate collisions. However, this can be understood from simple geometrical considerations on the interplay between the decrease of total spectator electric charge and that of the collision impact parameter $b$ with increasing centrality. Analogous EM-induced effects are present for collective azimuthal anisotropies (directed flow) for charged pion emission w.r.t.~the reaction plane as described in~\cite{zako2014}.

\begin{figure}[ht]
\centering
\includegraphics[width=13.5cm,clip]{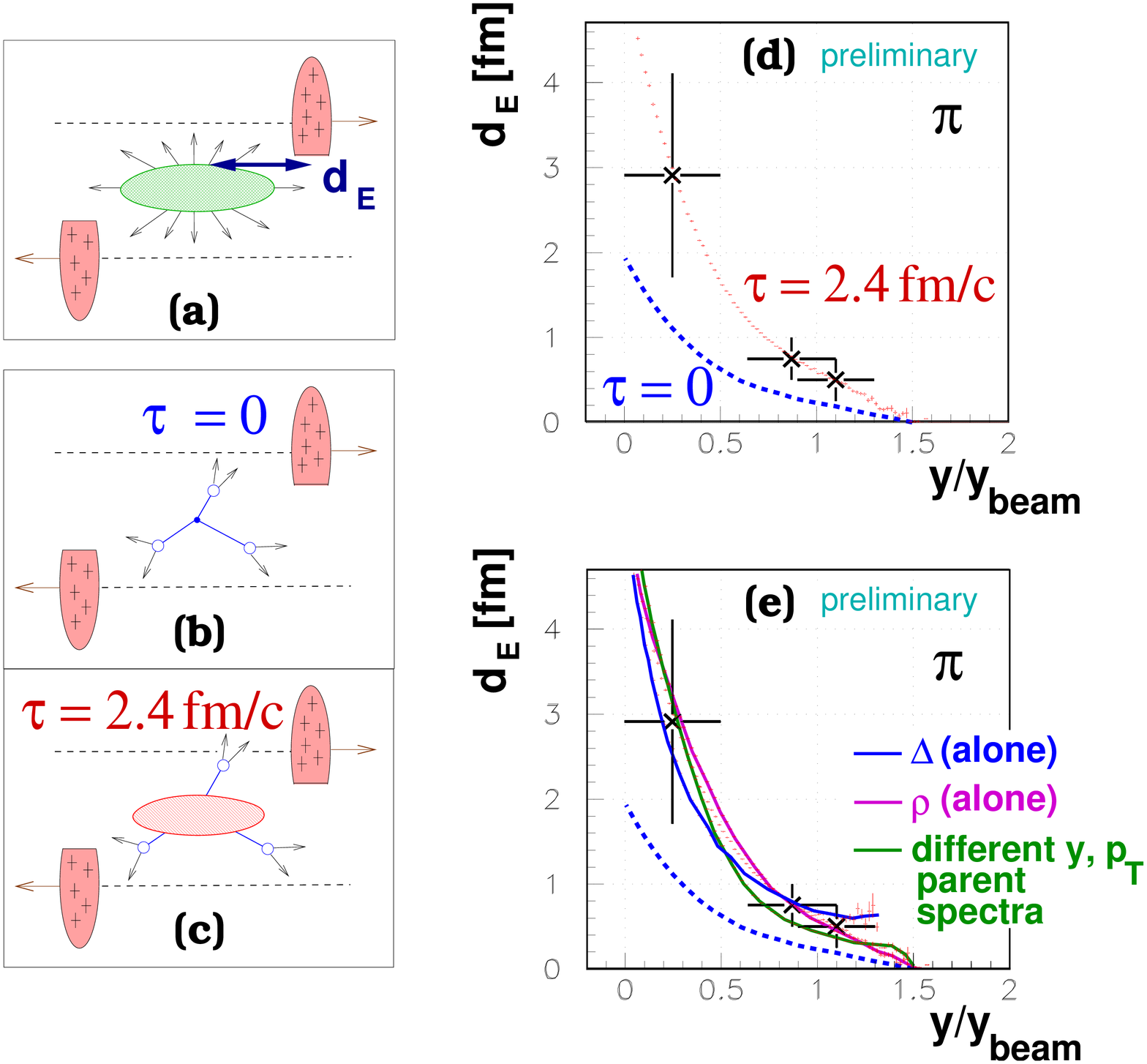}
  \caption{Information obtained on the basis of EM effects described in section~\ref{emeff}:
(a) schematic definition of $d_E$,
(b) simulation of resonances produced instantly at the moment of the collision ($\tau=0$) and (c) produced from an intermediate system characterized by a given proper lifetime $\tau$,
(d) dependence of $d_E$ on 
pion rapidity deduced from EM effects (data points), compared to the results of our Monte Carlo simulation illustrated in panels (b) and (c), respectively. 
(e) uncertainty of the present analysis induced by our incomplete knowledge of resonance production, estimated by 
assuming 
different characteristics for the input mix of parent $\Delta,\rho$ resonances.
All the panels originate from~\cite{wpcf2015}.}
\label{fig-1x}       
\end{figure}

\section{Sensitivity of EM effects to the space-time evolution of the system}

As the electromagnetically modified trajectory connects the charged meson produced at freeze-out with its final state measured in the detector, it is probably not very surprising that the EM-induced distortion of the latter appears sensitive to the space-time evolution of the particle production process. As this was demonstrated in~\cite{twospec,twospec-v1}, the overall shape of the electromagnetic distortion of charged pion spectra, and of charge-dependent effects on directed flow, is directly sensitive to the distance $d_E$ between the longitudinal 
position of the pion created at freeze-out, and the longitudinal 
position of the spectator system at the moment (in the collision c.m.s.~time) at which the pion is created 
as shown in Fig.~\ref{fig-1x} (a). 
Our 
subsequent studies based on a simulation of the relativistic motion of charged particles in the EM field indicate that the distortion of $\pi^+$ and $\pi^-$ spectra shown in Fig.~\ref{fig-1} corresponds to an average longitudinal position of the formation zone of fast ($y\approx y_{beam}$) pions being about $d_E\approx 0.75$~fm behind the spectator system. Simultaneously, the charge splitting of directed flow seen by the STAR Collaboration for Au+Au collisions at $\sqrt{s_{NN}}=7.7$~GeV~\cite{star2014} would correspond to $d_E\approx 3$~fm for pions produced at low c.m.s.~rapidities, while the very large values of $\pi^+$ directed flow seen in target vicinity in top SPS energy Pb+Pb collisions by the WA98 experiment~\cite{wa98} would suggest $d_E\approx 0.5$~fm. 
The situation is summarized in Fig.~\ref{fig-1x} (d), where a decrease of $d_E$ with increasing pion rapidity is evident. 
As we described in~\cite{wpcf2015}, a Monte Carlo model based on a compilation of knowledge existing on baryonic ($\Delta$) and mesonic ($\rho$) resonance production and decay can  serve as a simple tool for a further investigation of this result. A tentative simulation assuming {\em immediate}
production of these low-lying resonances at the interaction point --  
$\tau=0$, as shown in Fig.~\ref{fig-1x}(b) --
fails to reproduce the values of $d_E$ obtained from EM effects. This is no longer the case for the space-time evolution simulated as shown in Fig.~\ref{fig-1x}(c), which indicates that the space-time properties and proper lifetime $\tau$ of the intermediate system of hot and dense matter created in the collision can be independently studied, and estimated, by means of spectator-induced EM interactions. A tentative evaluation of uncertainties induced by our incomplete knowledge of resonance production is presented in Fig.~\ref{fig-1x}(e). This suggests that the latter uncertainties will in fact remain below the tentative error bars associated at present to our $d_E$ extraction procedures mentioned above. As the latter can definitely be reduced by future experimental and theoretical effort, it can be expected that an overall improvement in knowledge of space-time evolution of the system up to very high rapidities can be obtained from EM effects. 

At the present moment, the space-time 
characteristics of this system, tentatively
established from these
effects in the collision energy range $\sqrt{s_{NN}}=7.7-17.3$~GeV discussed here
and shown in
Figs~\ref{fig-1x}(d) and~\ref{fig-1x}(c), 
can give an independent estimate for the time of pion emission as a function of its rapidity.
At midrapidity this time being equal to $d_E(y/y_\mathrm{beam}=0)/\beta$, where $\beta$ is the spectator velocity, our preliminary estimate gives $5.3\pm 2.2$~fm/c for the time of pion emission in this region. For comparison, the pion decoupling times compiled
in the published HBT analysis by the ALICE Collaboration~\cite{alicehbt}
show values of approximately $\tau_f\approx 6$~fm/c
 for the same range of collision energies\footnote{The corresponding values of $\tau_f$ were directly read off the plot published in~\cite{alicehbt}.}.

%


\section{Conclusions}

We conclude at present that spectator-induced electromagnetic effects are a promising field which can be used to obtain new information of the space-time evolution of the system created in heavy ion collisions. 
There are good reasons to believe that the applicability of these effects for the latter purpose 
would extend over a broad range of centralities and not remain confined to peripheral reactions, as shown in Fig.~\ref{fig-1}(c).
A further development of these studies, including in particular new measurements at SPS energies~\cite{spscAddendumPbPb}, will reinforce the present relatively limited database on these phenomena at high energies. Consequently, this can be a chance to significantly deepen our present understanding of the space-time evolution and longitudinal expansion of hot and dense matter created in the course of the collision.\\

\begin{acknowledgement}
The authors warmly thank the organizers of the MESON 2016 conference, Krak\'ow, Poland.
This work was supported by the National Science Centre, Poland
(grant no. 2014/14/E/ST2/00018).
\end{acknowledgement}
%

\begin{thebibliography}{00}

\bibitem{zako2014}
  A.~Rybicki, A.~Szczurek and M.~K\l{}usek-Gawenda,
  {\it Acta Phys.\ Polon.} {\bf B46}, 
  no. 3, 737 (2015), and references therein.

 \bibitem{twospec}
  A.~Rybicki, A.~Szczurek, {\it Phys. Rev.} {\bf C75}, 054903 
(2007).

\bibitem{twospec-v1} 
  A.~Rybicki and A.~Szczurek,
  {\it Phys. Rev.} {\bf C87}, 054909 (2013).


 %

 %

\bibitem{wpcf2015}
  A.~Rybicki, A.~Szczurek, M.~Kie\l{}bowicz, N.~Davis and V.~Ozvenchuk,
  arXiv:1603.07558 [nucl-th], to appear in {\it Acta Phys.\ Polon.\ Supp.\,} and references therein.


\bibitem{star2014} 
  STAR Collab., L.~Adamczyk {\it et al.}, 
  {\it Phys. Rev. Lett.}  {\bf 112}, 162301 (2014).

\bibitem{wa98}   %
  WA98 Collab., H.~Schlagheck, 
  {\it Nucl. Phys.} {\bf A663}, 725 (2000).




\bibitem{alicehbt} 
ALICE Collab., K.~Aamodt {\it et al.}, {\it Phys. Lett.} {\bf B696}, 328 
(2011).
 %

\bibitem{spscAddendumPbPb}
NA61 Collab.,
  N.~Abgrall {\it et al.},
CERN-SPSC-2015-038; SPSC-P-330-ADD-8.

\end{thebibliography}
%
%
%
%

\end{document}